\documentclass[12pt]{article}
\usepackage{epsfig}
\usepackage{rotate}

\newcommand{\gsim}{~{}_{\textstyle\sim}^{\textstyle >}~}
\newcommand{\lsim}{~{}_{\textstyle\sim}^{\textstyle <}~}

\def\kpnn{K^+\rightarrow\pi^+\nu\overline{\nu}}

\def\klnn{K_{\rm L}\rightarrow\pi^0\nu\overline{\nu}}

\newcommand{\imlt}{\IM\lambda_t}
\newcommand{\relt}{\RE\lambda_t}
\newcommand{\relc}{\RE\lambda_c}

\newcommand{\bea}{\begin{eqnarray}}
\newcommand{\eea}{\end{eqnarray}}
\newcommand{\bd}{\begin{displaymath}}
\newcommand{\ed}{\end{displaymath}}
\newcommand{\be}{\begin{equation}}
\newcommand{\ee}{\end{equation}}

\newcommand{\gev}{\, {\rm GeV}}
\newcommand{\mev}{\, {\rm MeV}}

\def\bbuildrel#1_#2^#3{\mathrel{\mathop{\kern 0pt#1}\limits_{#2}^{#3}}}
\def\slash#1{\setbox0=\hbox{$#1$}#1\hskip-\wd0\dimen0=5pt\advance
       \dimen0 by-\ht0\advance\dimen0 by\dp0\lower0.5\dimen0\hbox
         to\wd0{\hss\sl/\/\hss}}

\newcommand{\RE}{{\rm Re}}
\newcommand{\IM}{{\rm Im}}

\newcommand{\vcb}{|V_{cb}|}

\newcommand{\vub}{|V_{ub}/V_{cb}|}

\renewcommand{\thefootnote}{\fnsymbol{footnote}}

\newcommand{\beq}{\begin{equation}}
\newcommand{\eeq}{\end{equation}}

\def\con{\ifmmode \hbox{\bf*} \else{\bf*}\fi}   
\def\scon{\ifmmode \hbox{\footnotesize\rm\bf*} \else{\footnotesize\rm\bf*}\fi}

\def\0#1{\relax\ifmmode\mathaccent"7017{#1}
        \else\accent23#1\relax\fi}              

\begin{document}

 
\begin{titlepage}
\vspace*{-1truecm} 
\begin{flushright} 
\begin{tabular}{l} 
DESY 01--054\\
TUM--HEP--412/01\\
hep--ph/0104238\\
August 2001
\end{tabular} 
\end{flushright} 
 
\vspace*{0.5truecm} 
 
\begin{center} 
\boldmath 
{\Large\bf Bounds on the Unitarity Triangle, 
$\sin 2\beta$}\\
\vspace*{0.3truecm}
{\Large\bf and $K\to\pi \nu\overline{\nu}$ Decays in Models with}\\
\vspace*{0.3truecm}
{\Large\bf  Minimal Flavour Violation} 
\unboldmath 
 
\vspace*{1.1cm} 
 
\smallskip 
\begin{center}
{\sc {\large Andrzej J. Buras}}\footnote{E-mail: {\tt  
Andrzej.Buras@ph.tum.de}} \\ 
\vspace*{2mm} 
{\sl Technische Universit\"at M\"unchen, 
Physik Department, D--85748 Garching, Germany}
\vspace*{1truecm}\\ 
{\sc {\large Robert Fleischer}}\footnote{E-mail: {\tt  
Robert.Fleischer@desy.de}} \\
\vspace*{2mm} 
{\sl Deutsches Elektronen-Synchrotron DESY, Notkestra\ss e 85,  
D--22607 Hamburg, Germany}
\end{center} 
 
\vspace{1.1truecm} 
 
{\large\bf Abstract\\[10pt]} \parbox[t]{\textwidth}{ 
We present a general discussion of the unitarity triangle from 
$\varepsilon_K$, $\Delta M_{d,s}$ and $K \to \pi\nu\overline{\nu}$ 
in models with minimal flavour violation (MFV), allowing for arbitrary 
signs of the generalized Inami--Lim functions $F_{tt}$ and $X$ relevant 
for $(\varepsilon_K,\Delta M_{d,s})$ and $K \to \pi\nu\overline{\nu}$, 
respectively. In the models in which $F_{tt}$ has a sign opposite to the 
one in the Standard Model, i.e.\ $F_{tt}<0$, the data for 
$(\varepsilon_K, \Delta M_{d,s})$ imply an absolute lower bound on the 
$B_d\to\psi K_{\rm S}$ CP asymmetry $a_{\psi K_{\rm S}}$
of 0.69, which is substantially stronger than 0.42 arising in the case of 
$F_{tt}>0$. An important finding of this paper is the observation 
that for given $Br(K^+\to\pi^+\nu\overline{\nu})$ and $a_{\psi K_{\rm S}}$ 
only {\it two} values for $Br(K_{\rm L}\to\pi^0\nu\overline{\nu})$, 
corresponding to the two signs of $X$, are possible in the full class of MFV 
models, independently of any new parameters arising in these models. This 
provides a powerful test for this class of models. Moreover, we derive 
{\it absolute} lower and upper bounds on 
$Br(K_{\rm L}\to\pi^0\nu\overline{\nu})$ as functions of 
$Br(K^+\to\pi^+\nu\overline{\nu})$. Using the present experimental upper 
bounds on $Br(K^+\to\pi^+\nu\overline{\nu})$ and $|V_{ub}/V_{cb}|$, we
obtain the absolute upper bound $Br(K_{\rm L}\to\pi^0\nu\overline{\nu})< 7.1
\cdot 10^{-10}$ \mbox{($90\%$ C.L.).}} 
 
\vskip1.5cm 
 
\end{center} 
 
\end{titlepage}

\thispagestyle{empty} 
\vbox{} 
\newpage 
  
\setcounter{page}{1}

\setcounter{footnote}{0} 
\renewcommand{\thefootnote}{\arabic{footnote}}

%
%
\section{Introduction}
The exploration of CP violation in $B_d \to \psi K_{\rm S}$ decays and 
the related determination of the angle $\beta$ in the usual unitarity 
triangle of the Cabibbo--Kobayashi--Maskawa (CKM) matrix are hot topics 
in present particle physics \cite{BaBar}-\cite{BePe}. The corresponding 
time-dependent CP asymmetry takes the following general form:
\begin{equation}\label{time-dep}
a_{\psi K_{\rm S}}(t)\equiv\frac{\Gamma(B^0_d(t)\to \psi K_{\rm S})-
\Gamma(\overline{B^0_d}(t)\to \psi K_{\rm S})}{\Gamma(B^0_d(t)\to 
\psi K_{\rm S})+\Gamma(\overline{B^0_d}(t)\to \psi K_{\rm S})}
={\cal A}_{\rm CP}^{{\rm dir}}\,\cos(\Delta M_d t)+
{\cal A}_{\rm CP}^{\rm mix}\,\sin(\Delta M_d t),
\end{equation}
where the rates correspond to decays of initially, i.e.\ at time $t=0$, 
present $B^0_d$- or $\overline{B^0_d}$-mesons, and $\Delta M_d>0$ 
denotes the mass difference between the mass eigenstates of the 
$B^0_d$--$\overline{B^0_d}$ system. The quantities 
${\cal A}_{\rm CP}^{{\rm dir}}$ and ${\cal A}_{\rm CP}^{\rm mix}$
are usually referred to as ``direct'' and ``mixing-induced'' CP-violating 
observables, respectively. In the Standard Model (SM), (\ref{time-dep}) 
simplifies as follows \cite{Bigi}:
\be\label{a1}
a_{\psi K_{\rm S}}(t)=-\sin 2\beta \sin(\Delta M_d t)\equiv 
-\,a_{\psi K_{\rm S}}\sin(\Delta M_d t),
\ee
thereby allowing the extraction of $\sin 2\beta$. It should be 
noted that a measurement of a non-vanishing value of 
${\cal A}_{\rm CP}^{{\rm dir}}$ at the level of 10\% would be a 
striking indication for new physics, as emphasized in a recent 
analysis of the $B\to \psi K$ system \cite{FM01}. However, for the
particular kind of physics beyond the SM considered in the present 
paper, direct CP violation in $B_d\to\psi K_{\rm S}$ decays 
is negligible.

In the future, $\sin 2\beta$ can also be determined through the 
measurement of the branching ratios for the rare decays $\kpnn$ 
and $\klnn$ \cite{BBSIN}. In the SM, we have to an excellent 
approximation
\begin{equation}\label{sin}
\sin 2\beta=\frac{2 r_s}{1+r^2_s},
\end{equation}
with
\begin{equation}\label{cbb}
r_s=\sqrt{\sigma}{\sqrt{\sigma(B_1-B_2)}-P_c(\nu\overline{\nu})
\over\sqrt{B_2}}\,.
\end{equation}
Here $B_1$ and $B_2$ are the following ``reduced'' branching ratios:
\begin{equation}\label{b1b2}
B_1={Br(\kpnn)\over 4.42\cdot 10^{-11}},\qquad
B_2={Br(\klnn)\over 1.93\cdot 10^{-10}},
\end{equation}
the quantity $P_c(\nu\overline{\nu})=0.40\pm0.06$ \cite{BB98} describes 
the internal charm-quark contribution to $\kpnn$, and 
\begin{equation}\label{sigma-def}
\sigma\equiv \frac{1}{(1-\lambda^2/2)^2},
\end{equation} 
with $\lambda$ being one of the Wolfenstein parameters \cite{WO}. 
In writing (\ref{sin}), we have assumed that $\sin 2\beta>0$, 
as expected in the SM. The numerical values in (\ref{b1b2}) and the
value for $P_c(\nu\overline{\nu})$ differ slightly from those given in 
\cite{BBSIN,BB98} due to $\lambda=0.222$ used here instead of $\lambda=0.22$
used in these papers. We will return to this point below. 

The strength of formulae (\ref{a1}) and (\ref{sin}) is their theoretical 
cleanness, allowing a precise determination of $\sin 2\beta$ free of 
hadronic uncertainties that is independent of other parameters like 
$\vcb$, $\vub$ and $m_t$. Therefore the comparison of these two determinations 
of $\sin 2\beta$ with each other is particularly well suited for tests of 
CP violation in the SM, and offers a powerful tool to probe the physics 
beyond it \cite{BBSIN,NIR}.

The simplest class of extensions of the SM are those models with ``minimal 
flavour violation'' (MFV) in which the contributions of any new operators 
beyond those present in the SM are negligible. In these models, all 
flavour-changing transitions are still governed by the CKM matrix, with 
no new complex phases beyond the CKM phase \cite{CDGG,UUT}.
If one assumes, in addition, that all new-physics contributions which 
are not proportional to $V_{td(s)}$ are negligible \cite{UUT}, then 
all the SM expressions for the decay amplitudes and particle--antiparticle 
mixing can be generalized to the MFV models by simply replacing the 
$m_t$-dependent Inami--Lim functions \cite{IL} by the corresponding 
functions $F_i$ in the extensions of the SM. The latter functions 
acquire now additional dependences on the parameters present in these 
extensions. Examples are the Two-Higgs-Doublet Model II (THDM) and 
the constrained MSSM if $\tan\bar\beta=v_2/v_1$ is not too large.
For MFV models, direct CP violation in $B_d\to\psi K_{\rm S}$ is 
negligible and the $\cos(\Delta M_d t)$ term in (\ref{time-dep}) 
vanishes.

Let us consider the off-diagonal element of the 
$B^0_{q}$--$\overline{B^0_{q}}$ mixing matrix as an example ($q\in\{d,s\}$). 
In the SM, we have (for a detailed discussion, see \cite{BF-rev})
\begin{equation}\label{M12}
M_{12}^{(q)}=\frac{G_{\rm F}^2M_W^2}{12\pi^2}\eta_Bm_{B_q}
\hat B_{B_q}F_{B_q}^2(V_{tq}^\ast V_{tb})^2 
S_0(x_t)\,e^{i(\pi-\phi_{\rm CP}(B_q))},
\end{equation}
where $\hat B_{B_q}$ is a non-perturbative parameter, $F_{B_q}$ the
$B_q$-meson decay constant, and $\eta_B=0.55$ a perturbative 
QCD factor \cite{BJW90,UKJS}, which is common to $M_{12}^{(d)}$ and 
$M_{12}^{(s)}$. Finally, the convention-dependent phase $\phi_{\rm CP}(B_q)$ 
is defined through
\begin{equation}
({\cal CP})|B^0_q\rangle=e^{i\phi_{\rm CP}(B_q)}|\overline{B^0_q}\rangle.
\end{equation}
In the MFV models, we have just to replace the Inami--Lim function 
$S_0(x_t)$ resulting from box diagrams with $(t,W^\pm)$ exchanges 
through an appropriate new function, which we denote by $F_{tt}$ 
\cite{ALI00,UUT}:
\begin{equation}\label{Ftt}
S_0(x_t)\to F_{tt}.
\end{equation}
Expression (\ref{M12}) plays a key role for (\ref{a1}), as 
$\Delta M_d=2|M_{12}^{(d)}|$, and $2\beta$ results from the difference
of $\mbox{arg}(M_{12}^{(d)})$ and the weak phase of the 
$B_d\to\psi K_{\rm S}$ decay amplitude, where the convention-dependent 
quantity $\phi_{\rm CP}(B_q)$ cancels. 

Two interesting properties of the MFV models have recently been pointed out 
\cite{UUT,ABRB}:
\begin{itemize}
\item
There exists a universal unitarity triangle (UUT) \cite{UUT} common to all 
these models and the SM that can be constructed by using measurable 
quantities that depend on the CKM parameters but are not polluted by the 
new parameters present in the extensions of the SM. These quantities simply 
do not depend on the functions $F_i$.
\item
There exists an absolute lower bound on $\sin 2\beta$ \cite{ABRB} that 
follows from the interplay of $\Delta M_d$ and $\varepsilon_K$, measuring
``indirect'' CP violation in the neutral kaon system. It depends only on 
$\vcb$ and $\vub$, as well as on the non-perturbative parameters 
$\hat B_K$, $F_{B_d}\sqrt{\hat B_d}$ and $\xi$ entering the standard 
analysis of the unitarity triangle.
\end{itemize}
The UUT can be constructed, for instance, by using $\sin 2\beta$ from 
(\ref{a1}) or (\ref{sin}), and the ratio $\Delta M_s/\Delta M_d$. 
The relevant formulae can be found in \cite{UUT}, where also other 
quantities suitable for the determination of the UUT are discussed. 
Concerning the lower bound on $\sin 2\beta$, a conservative 
scanning of all relevant input parameters gives \cite{ABRB,ERICE}
\be
\label{bound}
(\sin 2\beta)_{\rm min}=0.42,
\ee
corresponding to $\beta\geq 12^\circ$. This bound could be 
considerably improved when the values of $\vcb$, $\vub$, $\hat B_K$, 
$F_{B_d}\sqrt{\hat B_d}$, $\xi$ and -- in particular of 
$\Delta M_s$ -- will be known better \cite{ABRB,ERICE}. A handy approximate 
formula for $\sin 2\beta$ as a function of these parameters has recently
been given in \cite{BePe}. Using less conservative ranges of parameters,
these authors find $(\sin 2\beta)_{\rm min}=0.52$.

There is also an upper bound on $\sin 2\beta$, which is valid for the
Standard Model and the full class of MFV models. It is simply given by 
\cite{BLO}
\be
\label{ubound}
(\sin 2\beta)_{\rm max}=2 R_b^{\rm max}\sqrt{1-(R_b^{\rm max})^2}
\approx 0.82,
\ee
where
\begin{equation}\label{2.94}
R_b \equiv \frac{| V_{ud}^{}V^*_{ub}|}{| V_{cd}^{}V^*_{cb}|}
= \sqrt{\bar\varrho^2 +\bar\eta^2}
= \left(1-\frac{\lambda^2}{2}\right)\frac{1}{\lambda}
\left| \frac{V_{ub}}{V_{cb}} \right|
\end{equation}
is one side of the unitarity triangle.
Here \cite{BLO},
\be\label{gen-W}
\bar\varrho\equiv\varrho(1-\lambda^2/2), \qquad
\bar\eta\equiv\eta(1-\lambda^2/2),
\ee
where $\lambda$, $\varrho$ and $\eta$  are Wolfenstein parameters 
\cite{WO}. In obtaining the numerical value in (\ref{ubound}),
which corresponds to $\beta\lsim 28^\circ$, we have used 
$R_b^{\rm max}=0.46$.

In this paper, we would like to point out that the analyses of the MFV models 
performed in \cite{UUT,ABRB,ERICE,BePe} have implicitly assumed that the 
new functions $F_i$, summarizing the SM and new-physics contributions to 
$\varepsilon_K$, $\Delta M_{d,s}$ and $K\to\pi\nu\overline{\nu}$, have 
the same sign as the standard Inami--Lim functions. This assumption is 
certainly correct in the THDM and the MSSM. On the other hand, it cannot 
be excluded at present that there exist MFV models in which the functions 
$F_i$ relevant for $\varepsilon_K$, $\Delta M_s$ and 
$K\to\pi\nu\overline{\nu}$ have a sign {\it opposite} to the corresponding 
SM Inami--Lim functions. In fact, in the case of the $B\to X_s\gamma$ decay, 
such a situation is even possible in the MSSM if particular values of the 
supersymmetric parameters are chosen. Beyond MFV, scenarios in which the 
new-physics contributions to neutral meson mixing and rare $K$ decays 
were larger than the SM contributions and had opposite sign have been 
considered in \cite{BITALY}. Due to the presence of new complex phases 
in these general scenarios and new sources of flavour violation, the
predictive power of the corresponding models is much smaller than of the 
MFV models considered here.

In the following, we would like to generalize the existing formulae for
the MFV models to arbitrary signs of the generalized Inami--Lim functions 
$F_i$ and investigate the implications of the sign reversal in question 
for the determination of $\sin 2\beta$ and the unitarity triangle (UT) 
through $a_{\psi K_{\rm S}}$, $\varepsilon_K$, $\Delta M_{d,s}$ and 
$K\to\pi\nu\overline{\nu}$. In this context, we will also discuss strategies, 
allowing a direct determination of the sign of $F_{tt}$. However, the major 
findings of this paper deal with the rare kaon decays $\kpnn$ and $\klnn$. In 
particular, we point out that -- for given $Br(\kpnn)$ 
and $a_{\psi K_{\rm S}}$ -- only {\it two} values for $Br(\klnn)$, 
corresponding to the two possible signs of the generalized Inami--Lim 
function $X$, are possible in the full class of MFV models, independently 
of any new parameters present in these models. This feature provides an 
elegant strategy to check whether a MFV model is actually realized in nature 
and -- if so -- to determine the sign of $X$. Moreover, we derive 
{\it absolute} lower and upper bounds on the branching ratio $Br(\klnn)$ as 
a function of $Br(\kpnn)$, and emphasize the utility of 
$B\to X_s\nu\overline{\nu}$ decays to obtain further constraints. The
branching ratio $Br(\kpnn)$ and the CP asymmetry $a_{\psi K_{\rm S}}$ 
should be known rather accurately prior to the measurement of $Br(\klnn)$.

Our paper is organized as follows: in Section 2, we analyse the unitarity 
triangle and $\sin 2\beta$ using $\Delta M_{d,s}$, $\varepsilon_K$ and 
$a_{\psi K_{\rm S}}$. Section 3 is devoted to the $K\to\pi\nu\overline{\nu}$
decays, and our conclusions are summarized in Section 4.

\section{\boldmath{$\sin 2\beta$} and the UT from \boldmath{$\Delta M_{d,s}$},
\boldmath{$\varepsilon_K$} and \boldmath{$a_{\psi K_{\rm S}}$} }
\subsection{\boldmath{$\sin 2\beta$} from \boldmath{$\Delta M_{d,s}$} and
\boldmath{$\varepsilon_K$}}
In MFV models, the new-physics contributions to $\Delta M_{d,s}$ 
can be parametrized by a single function $F_{tt}$, as we have noted in
(\ref{Ftt}). The same ``universal'' function enters also the observable
$\varepsilon_K$ \cite{ALI00,ABRB,UUT}. In the SM, it reduces to the 
Inami--Lim function $S_0(x_t)\approx 2.38$. 

An important quantity for our discussion is
the length of one side of the unitarity triangle, $R_t$, defined by
\begin{equation}\label{2.95}
R_t \equiv \frac{| V_{td}^{}V^*_{tb}|}{| V_{cd}^{}V^*_{cb}|} =
 \sqrt{(1-\bar\varrho)^2 +\bar\eta^2}
=\frac{1}{\lambda} \left| \frac{V_{td}}{V_{cb}} \right|.
\end{equation}
From $\Delta M_d$ and 
$\Delta M_d/\Delta M_s$, one finds \cite{UUT,ABRB,ERICE}
\begin{equation}\label{RT}
R_t= 1.10\, \frac{ R_0}{A}\frac{1}{\sqrt{|F_{tt}|}} \quad\mbox{with}\quad
R_0\equiv\sqrt{\frac{\Delta M_d}{0.50/{\rm ps}}}
          \left[\frac{230\,\mbox{MeV}}{\sqrt{\hat B_d} F_{B_d}}\right]
          \sqrt{\frac{0.55}{\eta_B}}
\ee
and
\be\label{Rt}
R_t=0.83\, \xi\sqrt{\frac{\Delta M_d}{0.50/{\rm ps}}}
\sqrt{\frac{15.0/{\rm ps}}{\Delta M_s}} \quad\mbox{with}\quad
\xi \equiv
\frac{F_{B_s} \sqrt{\hat B_{B_s}}}{F_{B_d} \sqrt{\hat B_{B_d}}}\,,
\ee
respectively. 
The corresponding hadronic parameters were introduced
after (\ref{M12}). The Wolfenstein parameter $A$ is defined by 
$\vcb=A\lambda^2$.
These formulae show very clearly that the sign of 
$F_{tt}$ is immaterial for the analysis of $\Delta M_{d,s}$.

On the other hand, the  constraint from $\varepsilon_K$ reads \cite{ERICE}
\begin{equation}\label{100a}
\bar\eta \left[(1-\bar\varrho) A^2 \eta_2 F_{tt}
+ P_c(\varepsilon) \right] A^2 \hat B_K = 0.204\,,
\end{equation}
where $\eta_2=0.57$ is a perturbative QCD factor \cite{BJW90}, and 
$P_c(\varepsilon)=0.30\pm 0.05$ \cite{HN} summarizes the contributions 
not proportional to $V_{ts}^*V_{td}$.

Following \cite{ABRB}, but not assuming $F_{tt}$ to be positive, 
we find from (\ref{RT}) and (\ref{100a})
\be\label{main}
\sin 2\beta={\rm sgn}(F_{tt})\frac{1.65}{ R^2_0\eta_2}
\left[\frac{0.204}{A^2 B_K}
-\bar\eta P_c(\varepsilon)\right],
\ee
where the first term in the parenthesis is typically by a factor 2--3 
larger than the second term. We observe that the sign of $F_{tt}$ 
determines the sign of $\sin 2\beta$. Moreover, as (\ref{100a}) 
implies $\bar\eta<0$ for $F_{tt}<0$, also the sign of the second term 
in the parenthesis is changed. This means that, for a given set of input 
parameters, not only the sign of $\sin 2\beta$, but also its 
magnitude is affected by a reversal of the sign of $F_{tt}$.

At this point the following remark is in order. When using analytic 
formulae like (\ref{RT}), (\ref{Rt}) and (\ref{100a}) one should 
remember that the numerical constants given there are sensitive 
functions of $\lambda$. Consequently, varying $\lambda$ but keeping
these values fixed would result in errors. On the other hand, for 
fixed $\vcb$ any change of $\lambda$ modifies the parameter $A$ 
and consequently the impact of the variation of $\lambda$ within its
uncertainties on $\sin 2\beta$ and the unitarity triangle is very small.
The numerical values in (\ref{RT}), (\ref{Rt}) and (\ref{100a}) and the
value for $P_c(\varepsilon)$ differ slightly from those given in 
\cite{ABRB,ERICE} due to $\lambda=0.222$ used here instead of $\lambda=0.22$
used in these papers. Moreover, we have redefined $R_0$. 
This increase of $\lambda$ in question is made in order to be closer
to the experimental value of $|V_{ud}|$ \cite{SM}. 

The lower bound in (\ref{bound}) has been obtained by varying over all 
positive values of $F_{tt}$ consistent with the 
experimental values of $\Delta M_{d,s}$, $\vub$ and $\vcb$, and 
scanning all the relevant input parameters in the ranges given in 
Table~\ref{tab:inputparams}. Repeating this analysis for $F_{tt}<0$, 
we find
\be\label{10}
(-\sin 2\beta)_{{\rm min}}=0.69\,.
\ee
This result is rather sensitive to the minimal value of 
$\sqrt{\hat B_{B_d}}F_{B_d}$. Taking 
$(\sqrt{\hat B_{B_d}}F_{B_d})_{\rm min}=170\,\mbox{MeV}$ instead of 190\,MeV 
used in (\ref{10}), we obtain the bound of 0.51. For the same choice, 
the bound in (\ref{bound}) is decreased to 0.35. For 
$(\sqrt{\hat B_{B_d}}F_{B_d})_{\rm min}\ge 195\,\mbox{MeV}$ there are no 
solutions for $\sin 2\beta$ for the ranges of parameters given in 
Table~\ref{tab:inputparams}. Finally, only for $\hat B_K\ge 0.96$, 
$\vcb \ge 0.0414$ and $\vub\ge 0.094$ solutions for $\sin 2\beta$ 
exist. 

\begin{table}[t]
\vspace{0.4cm}
\begin{center}
\begin{tabular}{|c|c|c|}
\hline
{\bf Quantity} & {\bf Central} & {\bf Error}  
\\\hline
$\lambda$ & 0.222 &       \\
$|V_{cb}|$ & 0.041 & $\pm 0.002$      \\$\vub$ & $0.085$ & $\pm 0.018 $\\
$|V_{ub}|$ & $0.00349$ & $\pm 0.00076$ \\ 
$\hat B_K$ & 0.85 & $\pm 0.15$   \\
$\sqrt{\hat B_d} F_{B_{d}}$ & $230\mev$ & $\pm 40\mev$  \\
$m_t$ & $166\gev$ & $\pm 5\gev$   \\
$(\Delta M)_d$ & $0.487/\mbox{ps}$ & $\pm 0.014/\mbox{ps}$ \\
$(\Delta M)_s$  & $>15.0/\mbox{ps}$ & \\$\xi$ & $1.15$ & $\pm 0.06$  
\\
\hline
\end{tabular}
\caption[]{The ranges of the input parameters.\label{tab:inputparams}}
\end{center}
\end{table}

We conclude that in the case of $F_{tt}<0$ the lower bound on 
$|\sin 2\beta|$ is substantially stronger than for a positve $F_{tt}$. 
This is not surprising because in this case the contributions to 
$\varepsilon_K$ proportional to $V_{ts}^*V_{td}$ interfere destructively 
with the charm contribution. Consequently, $|\sin 2\beta|$ has to be larger 
to fit $\varepsilon_K$. Our discussion also shows that the decrease in the 
uncertainties of the parameters in Table~\ref{tab:inputparams} could 
well soon exclude all MFV models with $F_{tt}<0$.

\subsection{\boldmath{$a_{\psi K_{\rm S}}$}}
Concerning $a_{\psi K_{\rm S}}$, the situation is a bit more involved. 
As we have noted after (\ref{Ftt}), the angle $2\beta$ in (\ref{a1}) 
originates from 
\begin{equation}
2\beta=\mbox{arg}(M_{12}^{(d)})-\phi_{\rm D}(B_d\to \psi K_{\rm S}),
\end{equation}
where $\phi_{\rm D}(B_d\to \psi K_{\rm S})$ denotes a characteristic
weak phase of the $B_d\to \psi K_{\rm S}$ amplitude. In the SM
expression (\ref{a1}), it has been taken into account that $S_0(x_t)>0$,
and it has been assumed implicitly that the bag parameter $\hat B_{B_d}$ 
is positive. As emphasized in \cite{GKN}, for $\hat B_{B_d}<0$, the sign 
in (\ref{a1}) would flip. However, this case appears very unlikely to us. 
Indeed, all existing non-perturbative methods give $\hat B_{B_d}>0$, which 
we shall also assume in our analysis. A similar comment applies to 
$\hat B_K$. However, since $S_0(x_t)$ is replaced by the new parameter 
$F_{tt}$ in the case of the MFV models, which needs not be positive, 
the following phase $\phi_d$ is actually probed by the CP asymmetry of
$B_d\to \psi K_{\rm S}$:
\begin{equation}\label{phid-def}
\phi_d=2\beta+\mbox{arg}(F_{tt}).
\end{equation}
Consequently, formula (\ref{a1}) is generalized as follows:
\be\label{CP-mod}
a_{\psi K_{\rm S}}=\sin\phi_d={\rm sgn}(F_{tt})\sin2\beta.
\ee
On the other hand, if we use (\ref{main}) to predict $a_{\psi K_{\rm S}}$,
the sign of the resulting CP asymmetry is unaffected:
\be
a_{\psi K_{\rm S}}=\frac{1.65}{ R^2_0\eta_2}
\left[\frac{0.204}{A^2 B_K}
-\bar\eta P_c(\varepsilon)\right].
\ee
However, its absolute value will generally be larger for $F_{tt}<0$.

This analysis demonstrates that in the MFV models $\sin 2\beta$ can either 
be positive, as in the SM, or negative. This implies that, in addition 
to the universal unitarity triangle proposed in \cite{UUT}, there exists 
another universal unitarity triangle with $\sin 2\beta<0$, which is valid 
for MFV models with $F_{tt}<0$. This also means that the ``true'' CKM 
angle $\beta$ in the MFV models can only be determined from 
$a_{\psi K_{\rm S}}$  and $\Delta M_s/\Delta M_d$ up to a sign that 
depends on the sign of $F_{tt}$. In the spirit of \cite{UUT}, one can 
distinguish these two cases by studying simultaneously $\varepsilon_K$ and 
$\Delta M_d$. If the data on $a_{\psi K_{\rm S}}$ should 
violate the bound in (\ref{10}) but satisfy (\ref{bound}), the full class 
of MFV models with $F_{tt}<0$ would be excluded by the measurement of 
$a_{\psi K_{\rm S}}(t)$ alone. If also the bound (\ref{bound}) should be 
violated, all MFV models would be excluded. The present experimental 
situation is given as follows:
\begin{equation}\label{B-factory}
a_{\psi K_{\rm S}}=\left\{\begin{array}{ll}
0.59\pm0.14\pm0.05&\mbox{(BaBar \cite{BaBar})}\\
0.99\pm0.14\pm0.06&\mbox{(Belle \cite{Belle})}\\
0.79^{+0.41}_{-0.44}&\mbox{(CDF \cite{CDF}).}
\end{array}\right.
\end{equation}
Combining these results with the earlier measurement by  
ALEPH $(0.84^{+0.82}_{-1.04}\pm 0.16)$ \cite{ALEPH}
gives the grand average 
\be
a_{\psi K_{\rm S}}=0.79\pm 0.10,
\label{ga}
\ee
which does not yet allow us to draw any definite conclusions. In particular,
the most recent $B$-factory results in (\ref{B-factory}) are no longer
in favour of a small value of $a_{\psi K_{\rm S}}$, so that not even 
the case corresponding to negative $F_{tt}$ can be excluded. On the other
hand, in view of the Belle result \cite{Belle}, the upper bound given in 
(\ref{ubound}) may play an important role to search for new physics in the 
future. We observe that whereas the BaBar result \cite{BaBar} is fully 
consistent with $\left|\sin2\beta\right|_{\rm max}=0.82$, corresponding 
to $|V_{ub}/V_{cb}|_{\rm max}=0.105$, the Belle result violates this bound. 
This can also be seen in Fig.~\ref{fig:s2bmax}, where we show 
$\left|\sin 2\beta\right|_{\rm max}$ as a function of 
$|V_{ub}/V_{cb}|_{\rm max}$.  
Only for values of $|V_{ub}/V_{cb}|$ that are substantially higher than 
the ones given in Table~\ref{tab:inputparams} could the Belle result be 
valid within the MFV models.
Finally, as seen from (\ref{10}) and Fig.~\ref{fig:s2bmax}, a decrease of 
$\vub_{\rm max}$ down to $0.085$ would put the MFV models with $F_{tt}<0$ 
into difficulties, independently of other input parameters in 
Table~\ref{tab:inputparams}.

\begin{figure}
\centerline{\rotate[r]{
\epsfysize=11.3truecm
{\epsffile{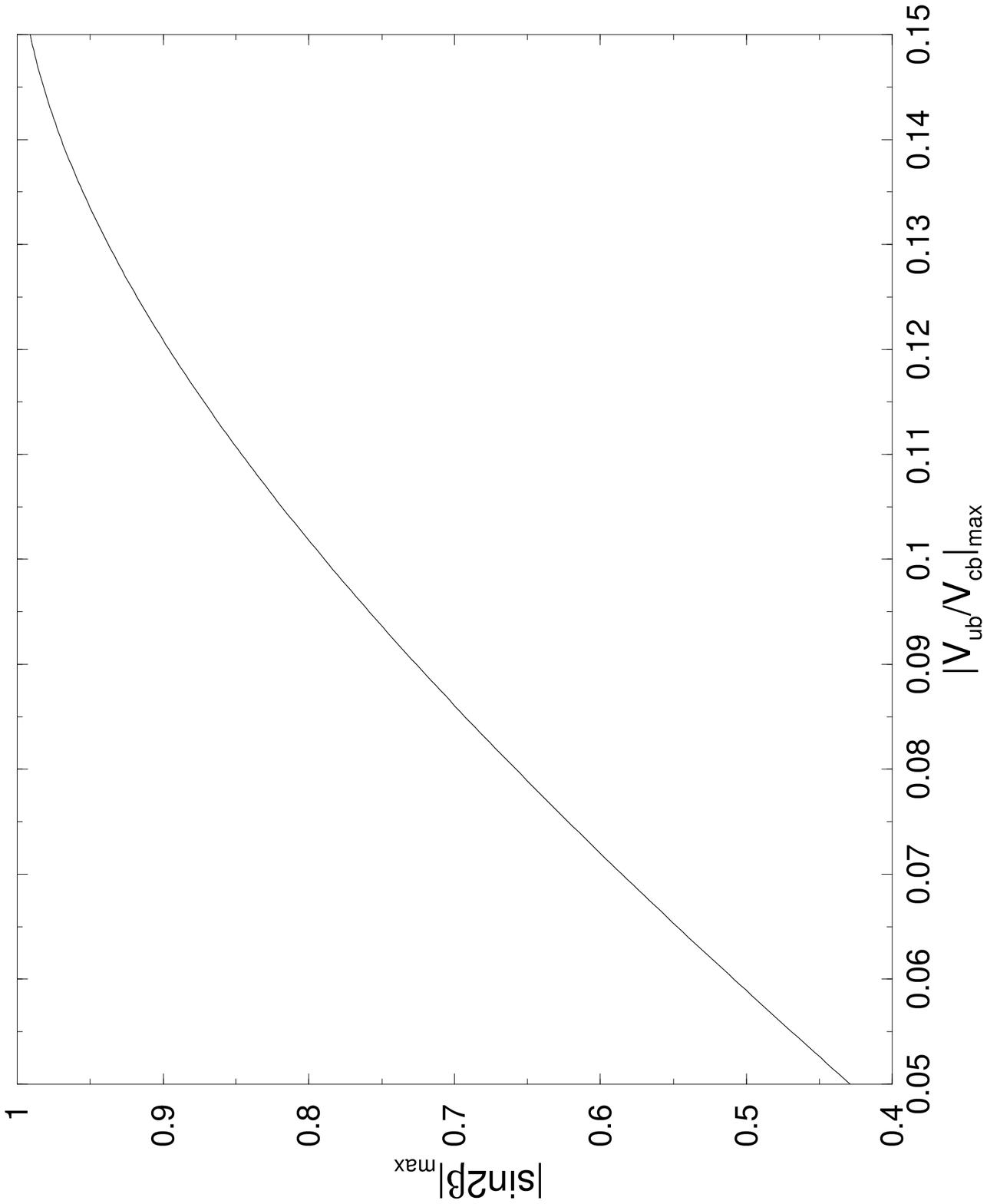}}}}
\caption{$\left|\sin 2\beta\right|_{\rm max}$ as a function of 
$|V_{ub}/V_{cb}|_{\rm max}$.}\label{fig:s2bmax}
\end{figure}

\subsection{Direct Determination of \boldmath{$\mbox{sgn}(F_{tt})$}}
It would of course be important to measure the sign of the parameter
$F_{tt}$ directly and to check the consistency with the bounds discussed
above. Several strategies were proposed to extract the phase 
$\phi_d$ introduced in (\ref{phid-def}) unambiguously \cite{ambig}.
This information would be very useful to distinguish between 
$F_{tt}>0$ and $F_{tt}<0$. Let us illustrate this by considering
an example, where we assume that $a_{\psi K_{\rm S}}=0.75$ has been 
measured, corresponding to $\phi_d=48.6^\circ$ or $131.4^\circ$. 
The strategies for the distinction between these two possibilities 
are discussed in the next paragraph. Let us then assume that the 
unambiguous determination of $\phi_d$ gives $48.6^\circ$. For 
$\mbox{arg}(F_{tt})=0$, we would then obtain $\beta=24.3^\circ$ or 
$\beta=204.3^\circ$, where the latter solution would be excluded by 
the data on $|V_{ub}/V_{cb}|$, requiring 
$\sqrt{\bar\varrho^2+\bar\eta^2}\lsim0.5$ (see our discussion 
below (\ref{gen-W})). For $\mbox{arg}(F_{tt})=180^\circ$,
we would get $\beta=114.3^\circ$ or $\beta=294.3^\circ$, which would 
both be excluded by $|V_{ub}/V_{cb}|$. Consequently, we would 
conclude $\beta=24.3^\circ$ and $\mbox{arg}(F_{tt})=0$ in this case,
which could also accommodate the Standard Model. On the other hand, if 
$\phi_d$ is found to be $131.4^\circ$, the situation is as follows: 
for $\mbox{arg}(F_{tt})=0$, we would get $\beta=65.7^\circ$ or 
$\beta=245.7^\circ$, which would both be excluded
by $|V_{ub}/V_{cb}|$. In the case of $\mbox{arg}(F_{tt})=180^\circ$,
we would obtain $\beta=-24.3^\circ$ or $\beta=155.7^\circ$, where the
latter solution would again be excluded by $|V_{ub}/V_{cb}|$. In 
this case, we would then conclude that $\beta=-24.3^\circ$ and 
$\mbox{arg}(F_{tt})=180^\circ$. Since the Standard Model cannot be
included in this category, we would have an unambiguous signal for new 
physics. 

The key element for the resolution of the twofold ambiguity in the extraction 
of $\phi_d$ from $a_{\psi K_{\rm S}}=\sin\phi_d$ is the determination 
of $\cos\phi_d$. For the example given in the previous paragraph, 
$\cos\phi_d=+0.66$ would imply MFV models with $F_{tt}>0$, containing
also the Standard Model, whereas $\cos\phi_d=-0.66$ would imply unambiguously
the presence of new physics, corresponding to $F_{tt}<0$ in MFV scenarios.
The quantity $\cos\phi_d$ can be probed through the angular distribution of 
$B_d\to\psi K^*[\to\pi^0K_{\rm S}]$ decays \cite{DDF}, allowing us to extract
\begin{equation}
\cos\delta_f\cos\phi_d.
\end{equation}
Here $\delta_f$ is a strong phase corresponding to a given final-state
configuration $f$ of the $\psi K^*$ system. Theoretical tools, such as
``factorization'', may be sufficiently accurate to determine 
$\mbox{sgn}(\cos\delta_f)$, thereby allowing the direct extraction of 
$\cos\phi_d$. In the case of $B_s$ decays, even information on the sign 
of $F_{tt}$ can be obtained in a direct way, as the SM ``background'' is 
negligibly small in 
\begin{equation}\label{phis-def}
\phi_s=-2\lambda^2\eta+\mbox{arg}(F_{tt})\approx \mbox{arg}(F_{tt}).
\end{equation}
In analogy to the $B_d\to\psi K^*[\to\pi^0K_{\rm S}]$ case, the quantity
\begin{equation}
\cos\tilde\delta_f\cos\phi_s=\cos\tilde\delta_f\,\mbox{sgn}(F_{tt})
\end{equation}
can be probed through the observables of the $B_s\to\psi \phi$
angular distribution \cite{DFN}. These modes are very accessible 
at hadron machines. Using again a theoretical input, such as
``factorization'', to determine $\mbox{sgn}(\cos\tilde\delta_f)$, 
the sign of $F_{tt}$ can be extracted. If $\phi_d$ is known unambiguously, 
$SU(3)$ flavour-symmetry arguments can be used to fix 
$\mbox{sgn}(\cos\tilde\delta_f)$ from $B_d\to\psi K^*$ decays \cite{DFN}; 
alternative ways to determine $\cos\phi_s=\mbox{sgn}(F_{tt})$ from $B_s$ 
decays were also noted in that paper.

\subsection{UUT from \boldmath{$a_{\psi K_{\rm S}}$} and 
\boldmath{$\Delta M_s/\Delta M_d$}}
In \cite{B95,UUT}, a construction of the UUT by means of $a_{\psi K_{\rm S}}$ 
and $R_t$ following from $\Delta M_s/\Delta M_d$ has been presented. 
Generally, for given values of $(a_{\psi K_{\rm S}},R_t)$,
there are eight solutions for $(\bar\varrho,\bar\eta)$. However, only two 
solutions are consistent with the bound in (\ref{ubound}), corresponding 
to the two possible signs of $F_{tt}$. 

For the derivation of explicit expressions for $\bar\varrho$ and $\bar\eta$, 
it is useful to consider
\begin{equation}\label{beta-def}
\mbox{sgn}(F_{tt})\,{\rm ctg}\beta=\frac{1-\bar\varrho}{|\bar\eta|}\equiv 
f(\beta),
\end{equation}
as (\ref{2.95}) implies
\begin{equation}
R_t^2=(1-\bar\varrho)^2+\bar\eta^2=\left[f(\beta)^2+1\right]\bar\eta^2.
\end{equation}
Consequently, admitting also negative $F_{tt}$, we obtain
\begin{equation}\label{r-e}
\bar\eta=\mbox{sgn}(F_{tt})\left[\frac{R_t}{\sqrt{f(\beta)^2+1}}\right],\quad
\bar\varrho=1-f(\beta)|\bar\eta|.
\end{equation}
If we take into account the constraint from $|V_{ub}/V_{cb}|$, yielding
$\bar\varrho<1$, we conclude that $f(\beta)$ is always positive. Moreover, 
as $a_{\psi K_{\rm S}}={\rm sgn}(F_{tt})\sin2\beta$, we may write
\begin{equation}\label{rs-beta2}
f(\beta)=\frac{1\pm\sqrt{1-a_{\psi K_{\rm S}}^2}}{a_{\psi K_{\rm S}}}=
{\rm sgn}(F_{tt})\left[\frac{1\pm|\cos2\beta|}{\sin2\beta}\right].
\end{equation}
Now the upper bound $|\beta|\lsim 28^\circ$ (see (\ref{ubound})) implies 
$|{\rm ctg}\beta|=f(\beta)\gsim1.9$. As $0<a_{\psi K_{\rm S}}<1$, 
the ``$-$'' solution in (\ref{rs-beta2}) is hence ruled out, and the 
measurement of $a_{\psi K_{\rm S}}$ determines $f(\beta)$ 
{\it unambiguously} through
\begin{equation}\label{rs-unambig}
f(\beta)=\frac{1+\sqrt{1-a_{\psi K_{\rm S}}^2}}{a_{\psi K_{\rm S}}}.
\end{equation}
Finally, with the help of (\ref{r-e}), we arrive at
\be\label{rho-eta}
\bar\eta={\rm sgn}(F_{tt})R_t
\sqrt{\frac{1-\sqrt{1-a_{\psi K_{\rm S}}^2}}{2}},\quad
\bar\varrho=1-\left[\frac{1+\sqrt{1-
a_{\psi K_{\rm S}}^2}}{a_{\psi K_{\rm S}}}\right]|\bar\eta|.
\ee
The function $f(\beta)$ plays also a key role for the analysis of the
$K\to\pi\nu\overline{\nu}$ system, which is the topic of Section 3.

\subsection{Lower and Upper Bounds on \boldmath{$J_{{\rm CP}}$} and 
\boldmath{$\imlt$}}
The areas $A_{\Delta}$ of all unitarity triangles are equal and related 
to the measure of CP violation $J_{\rm CP}$ \cite{CJ}:
\begin{equation}
\left|J_{\rm CP}\right| = 2 A_{\Delta}=
\lambda \left(1-\frac{\lambda^2}{2}\right)|\imlt|,
\end{equation}
where  $\lambda_t=V_{ts}^*V_{td}$. The cleanest measurement of 
$\imlt$ is offered by $Br(\klnn)$ \cite{BBSIN}, which is discussed in the
following section. The importance of the measurement of $J_{{\rm CP}}$ has 
been stressed in particular in \cite{Marciano}.

From $\varepsilon_K$ and $\Delta M_{d,s}$, we find the following 
absolute upper and lower bounds on $|\imlt|$ in the MFV models:
\be\label{uimlt}
|\imlt|_{\rm max}=\left\{\begin{array}{ll}
1.74\cdot 10^{-4}& F_{tt}>0  \\
 1.70\cdot 10^{-4}& F_{tt}<0 \\
\end{array}\right.
\end{equation}
and
\be\label{mimlt}
|\imlt|_{\rm min}=\left\{\begin{array}{ll}
0.55\cdot 10^{-4}& F_{tt}>0  \\
 1.13\cdot 10^{-4}& F_{tt}<0, \\
\end{array}\right.
\end{equation}
with ${\rm sgn}(\imlt)={\rm sgn}(F_{tt})$. In the SM, 
$0.94\cdot 10^{-4}\le \imlt\le 1.60\cdot 10^{-4}$, and the unitarity 
of the CKM matrix implies $|\imlt|_{\rm max}=1.83\cdot 10^{-4}$.

\section{\boldmath{$\sin 2\beta$} and UT from 
\boldmath{$K\to\pi\nu\overline{\nu}$}
in MFV Models}
\subsection{Preface}
In MFV models, the short-distance contributions to $\kpnn$ and 
$\klnn$ proportional to $V^*_{ts}V_{td}$ are described by a function 
$X$, resulting from $Z^0$ penguin and box diagrams. In evaluating 
$\sin 2\beta$ in terms of the branching ratios for $\kpnn$ and $\klnn$, 
the function $X$ drops out \cite{BBSIN}. Being determined from two branching 
ratios, there is a four-fold ambiguity in $\sin 2\beta$ that is reduced to a 
two-fold ambiguity if $\bar\varrho<1$, as required by the size of $\vub$. 
The left over solutions correspond to two signs of $\sin 2\beta$ that 
can be adjusted to agree with the analysis of $\varepsilon_K$. In 
the SM, the THDM and the MSSM, the functions $F_{tt}$ and $X$ are both 
positive, resulting in $\sin 2\beta$ given by (\ref{sin})--(\ref{b1b2}). 
We would now like to generalize this discussion and the SM formulae for 
$\kpnn$ and $\klnn$ to MFV models with arbitrary signs of $F_{tt}$ 
and $X$. As one of our major findings, we point out the interesting
feature that -- for given $Br(K^+\to\pi^+\nu\overline{\nu})$ and 
$a_{\psi K_{\rm S}}$ -- only {\it two} values for 
$Br(K_{\rm L}\to\pi^0\nu\overline{\nu})$, corresponding 
to the two signs of $X$, are possible in the full class of MFV models, 
independently of any new parameters arising in these models.

\subsection{\boldmath{$\kpnn$}}
The reduced branching ratio $B_1$ defined in (\ref{b1b2}) is given by
\begin{equation}\label{bkpn}
B_1=\left[{\imlt\over\lambda^5}|X|\right]^2+
\left[{\relc\over\lambda}{\rm sgn}(X) P_c(\nu\overline{\nu}) +
{\relt\over\lambda^5}|X|\right]^2,
\end{equation}
where $\lambda_t=V^\ast_{ts}V_{td}$ with
\be\label{RE-IM}
\imlt=\eta A^2 \lambda^5,\quad 
\relt=-\left(1-\frac{\lambda^2}{2}\right)A^2\lambda^5(1-\bar\varrho),
\ee
and $\lambda_c=-\lambda (1-\lambda^2/2)$. Therefore, the standard analysis 
of the unitarity triangle by means of $\kpnn$ \cite{BBSIN,BLO} can be 
generalized to arbitrary signs of $X$ and $F_{tt}$ through the replacements
\be
X\to |X|, \quad P_c(\nu\overline{\nu})\to {\rm sgn}(X) 
P_c(\nu\overline{\nu}),\quad \bar\eta \to \mbox{sgn}(F_{tt})|\bar\eta|.
\ee

We find then that the measured value of 
$Br(K^{+} \to \pi^{+} \nu \overline{\nu})$ 
determines an ellipse in the $(\bar\varrho,\bar\eta)$ plane,
\begin{equation}\label{ellipse}
\left(\frac{\bar\varrho-\varrho_0}{\bar\varrho_1}\right)^2+
\left(\frac{\bar\eta}{\bar\eta_1}\right)^2=1,
\end{equation}
centered at $(\varrho_0,0)$ with 
\begin{equation}\label{110}
\varrho_0 = 1 + {\rm sgn}(X) \frac{P_c(\nu\overline{\nu})}{A^2 |X|},
\end{equation}
and having the squared axes
\begin{equation}\label{110a}
\bar\varrho_1^2 = r^2_0, \quad \bar\eta_1^2 = \left( \frac{r_0}{\sigma}
\right)^2 \quad\mbox{with}\quad\,\,
r^2_0 = \frac{\sigma B_1}{A^4 |X|^2}\,.
\end{equation}
The ellipse (\ref{ellipse}) intersects with the circle (\ref{2.94}).
This allows us to determine $\bar\varrho$ and $\bar\eta$:
\begin{equation}\label{113}
\bar\varrho = \frac{1}{1-\sigma^2} \left[ \varrho_0 \mp \sqrt{\sigma^2
\varrho_0^2 +(1-\sigma^2)(r_0^2-\sigma^2 R_b^2)} \right], \quad
\bar\eta = {\rm sgn}(F_{tt})\sqrt{R_b^2 -\bar\varrho^2},
\end{equation}
and consequently
\begin{equation}\label{113aa}
R_t^2 = 1+R_b^2 - 2 \bar\varrho\,.
\end{equation}
Given $\bar\varrho$ and $\bar\eta$, one can determine $V_{td}$:
\begin{equation}\label{vtdrhoeta}
V_{td}=A \lambda^3(1-\bar\varrho-i\bar\eta),\quad
|V_{td}|=A \lambda^3 R_t.
\end{equation}

The deviation of $\varrho_0$ from unity measures the relative importance
of the internal charm contribution. For $X>0$, we have, as usual, 
$\varrho_0>1$ so that the ``$+$'' solution in (\ref{113}) is excluded 
because of $\varrho<1$. On the other hand, for $X<0$, the center of 
the ellipse is shifted to $\varrho_0<1$, and for 
$|X|\leq P_c(\nu\overline{\nu})/A^2$ can even be at $\varrho_0\leq0$.

\subsection{\boldmath{$\klnn$, $\kpnn$} and the Unitarity Triangle}
The reduced branching ratio $B_2$ defined in (\ref{b1b2}) is given by
\begin{equation}\label{bklpn}
B_2=\left[{\imlt\over\lambda^5}|X|\right]^2.
\end{equation}
Following \cite{BBSIN}, but admitting both signs of $X$ and $F_{tt}$, we find
\begin{equation}\label{rhetb}
\bar\varrho=1+\left[{\pm\sqrt{\sigma(B_1-B_2)}+
{\rm sgn}(X) P_c(\nu\overline{\nu})\over A^2 |X|}\right],\quad
\bar\eta= {\rm sgn}(F_{tt}){\sqrt{B_2}\over\sqrt{\sigma} A^2 |X|},
\end{equation}
where $\sigma$ was defined in (\ref{sigma-def}). Introducing
\begin{equation}\label{sinu}
r_s\equiv{1-\bar\varrho\over\bar\eta}={\rm ctg}\beta,
\end{equation}
we then find 
\begin{equation}\label{cbbnew}
r_s={\rm sgn}(F_{tt})\,\sqrt{\sigma}\left[{\mp\sqrt{\sigma(B_1-B_2)}-
{\rm sgn}(X)P_c(\nu\overline{\nu})\over\sqrt{B_2}}\right],
\end{equation}
with (\ref{sin}) and (\ref{b1b2}) unchanged. We observe that $r_s$ is 
independent of $|X|$ but the sign of the 
interference between the $V^*_{ts}V_{td}$  contribution and the charm 
contribution $P_c(\nu\overline{\nu})$ to $\kpnn$ matters.

In order to deal with the ambiguities present in (\ref{cbbnew}), we 
consider
\begin{equation}\label{EEE}
{\rm sgn}(F_{tt})\,r_s=\sqrt{\sigma}\left[{\mp\sqrt{\sigma(B_1-B_2)}-
{\rm sgn}(X)P_c(\nu\overline{\nu})\over\sqrt{B_2}}\right]=f(\beta),
\end{equation}
where $f(\beta)$ was introduced in (\ref{beta-def}). As we have noted 
after (\ref{r-e}), $f(\beta)$ has to be positive. Consequently, for 
$X>0$, only the ``$+$'' solution is allowed. On the other hand, in the 
case of $X<0$, the ``$-$'' solution gives also a positive value of 
$f(\beta)$ if 
\begin{equation}\label{B1-B2-limit}
B_1-B_2<\frac{P_c(\nu\overline{\nu})^2}{\sigma}\approx0.15.
\end{equation}
Numerical studies show that both $Br(\kpnn)$ and $Br(\klnn)$ have to 
be below $1 \cdot10^{-11}$ to satisfy (\ref{B1-B2-limit}). As such low 
values are extremely difficult to measure , we will not consider this 
possibility further, which leaves us with the ``$+$'' solution in
(\ref{cbbnew}).

In Table~\ref{ANAS}, we show the resulting values of 
$\mbox{sgn}(F_{tt})\sin 2\beta=a_{\psi K_{\rm S}}$ for several choices 
of $Br(\kpnn)$ and $Br(\klnn)$, setting $P_c(\nu\overline{\nu})=0.40$. 
We observe that the sign of $X$ is important; we also note that certain 
values violate the bounds in (\ref{bound}) and (\ref{ubound}). This implies 
that certain combinations of the two branching ratios are excluded within 
the MFV models. Let us then find out which combinations are still allowed.

\begin{table}[thb]
\caption[]{$\mbox{sgn}(F_{tt})\sin 2\beta=a_{\psi K_{\rm S}}$ 
in MFV models for specific values of $Br(\klnn)\equiv Br(K_{\rm L})$
and $Br(\kpnn)\equiv Br(K^+)$ for ${\rm sgn}(X)=+1~(-1)$ and 
$P_c(\nu\overline{\nu})=0.40$.
\label{ANAS}}
\begin{center}
\begin{tabular}{|c|c|c|c|}\hline
$Br(K_{\rm L})~[10^{-11}]$ & 
$ Br(K^+)=8.0~[10^{-11}]$ &  $ Br(K^+)=16~[10^{-11}]$ &
$ Br(K^+)=24~[10^{-11}]$  \\ \hline
2.0 & 0.60~(0.35) & 0.40~(0.27) & 0.31~(0.22) \\
3.0 & 0.71~(0.43) & 0.48~(0.32) & 0.38~(0.27)\\
4.0 & 0.79~(0.49) & 0.55~(0.37) & 0.43~(0.32)\\
5.0 & 0.86~(0.54) & 0.60~(0.42) & 0.48~(0.35)\\
6.0 & 0.91~(0.59) & 0.65~(0.45) & 0.52~(0.38)\\
7.0 & 0.94~(0.64) & 0.70~(0.49) & 0.56~(0.41)\\
8.0 & 0.97~(0.68) & 0.73~(0.52) & 0.60~(0.44)\\
 \hline
 \end{tabular}
\end{center}
\end{table}

\subsection{\boldmath{$Br(\klnn)$} from \boldmath{$a_{\psi K_{\rm S}}$}
 and \boldmath{$Br(\kpnn)$}}\label{sec:GOLD}
As $a_{\psi K_{\rm S}}$ and $Br(\kpnn)$ will be known rather accurately 
prior 
to the measurement of $Br(\klnn)$, it is of interest to calculate 
$Br(\klnn)$ as a function of $a_{\psi K_{\rm S}}$ and $Br(\kpnn)$. From 
(\ref{EEE}), we obtain
\be\label{B1B2}
B_1=B_2+\left[\frac{f(\beta)\sqrt{B_2}+
{\rm sgn}(X)\sqrt{\sigma}P_c(\nu\overline{\nu})}
{\sigma}\right]^2.
\ee
The important virtue of (\ref{B1B2}) when compared with (\ref{cbbnew})
is the absence of the ambiguity due to the $\mp$ in front of 
$\sqrt{\sigma(B_1-B_2)}$. 

As we have seen in (\ref{rs-unambig}), the measurement of 
$a_{\psi K_{\rm S}}$ determines $f(\beta)$ unambiguously.
This finding, in combination with (\ref{B1B2}), implies the 
following interesting property of the MFV models:
\begin{itemize}
\item
For given $a_{\psi K_{\rm S}}$ and $Br(\kpnn)$ only two values of 
$Br(\klnn)$, corresponding to the two possible signs of $X$, are possible 
in the full class of MFV models, independently of any new parameters 
present in these models. 
\end{itemize}
Consequently, measuring $Br(\klnn)$ will 
either select one of these two possible values or rule out all MFV models.
We would like to emphasize that the latter possibility could take place 
even if the lower bound on $|\sin 2\beta|$ \cite{ABRB} is satisfied by
the data on $a_{\psi K_{\rm S}}$, which is favoured by the most recent
$B$-factory results given in (\ref{B-factory}).

In Table~\ref{ANA}, we show values of $Br(\klnn)$ in the MFV models for 
specific values of $a_{\psi K_{\rm S}}$ and $Br(\kpnn)$ and the two signs 
of $X$. Note that the second column gives
the {\it absolute} lower bound on $Br(\klnn)$ in the MFV models as a function
of $Br(\kpnn)$. This bound follows simply from the lower bound in 
(\ref{bound}). On the other hand, the last column gives the corresponding
{\it absolute} upper bound. This bound is the consequence of the 
upper bound in (\ref{ubound}). The third column gives the 
lower bound on $Br(\klnn)$ corresponding to the bound in (\ref{10}) that
applies for a negative $F_{tt}$.

\begin{table}[t]
\caption[]{ Values of $Br(\klnn)$ in the
MFV models in units of $10^{-11}$ for specific values of $a_{\psi K_{\rm S}}$
and $Br(\kpnn)$ and ${\rm sgn}(X)=+1~(-1)$. We set 
$P_c(\nu\overline{\nu})=0.40$.
\label{ANA}}
\begin{center}
\begin{tabular}{|c|c|c|c|}\hline
$Br(\kpnn)$ $[10^{-11}]$ & 
$ a_{\psi K_{\rm S}}=0.42 $ &  $ a_{\psi K_{\rm S}}=0.69 $ &
$ a_{\psi K_{\rm S}}=0.82 $  \\ \hline
5.0 & 0.45~(2.0) & 1.4~(5.8) & 2.2~(8.6) \\
10.0 & 1.2~(3.5) & 3.8~(10.0) & 5.9~(15.0)\\
15.0 & 2.1~(4.8) & 6.3~(14.0) & 9.9~(21.1)\\
20.0 & 3.0~(6.2) & 9.0~(17.9) & 14.1~(27.0)\\
25.0 & 3.9~(7.5) & 11.8~(21.7) & 18.4~(32.8)\\
30.0 & 4.9~(8.7) & 14.6~(25.4) & 22.7~(38.6)\\
 \hline
 \end{tabular}
\end{center}
\end{table}

\begin{figure}
\centerline{\rotate[r]{
\epsfysize=10.3truecm
{\epsffile{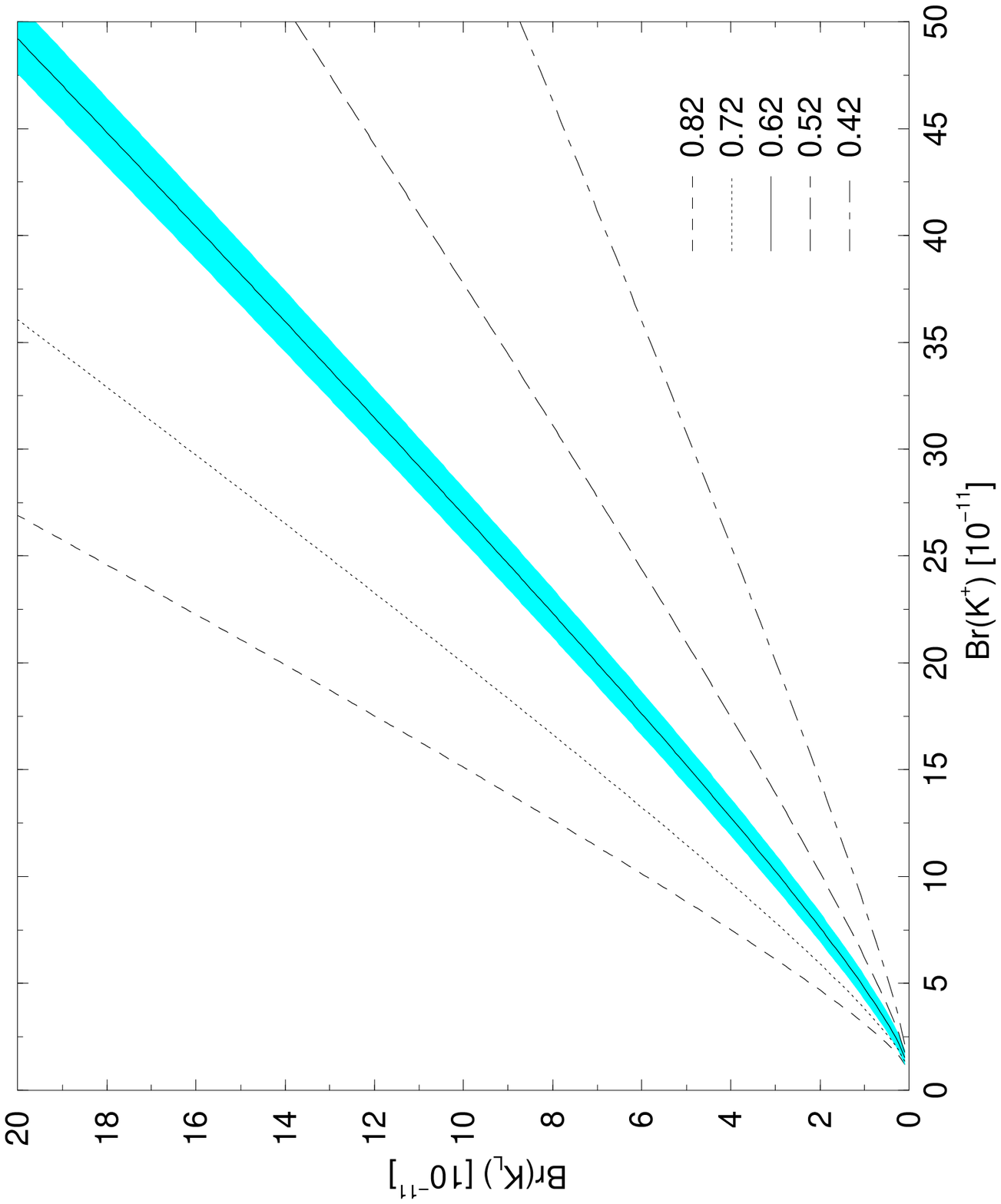}}}}
\caption{$Br(\klnn)$ as a function of $Br(\kpnn)$ for several 
values of $a_{\psi K_{\rm S}}$ in the case of ${\rm sgn}(X)=+1$.
For $a_{\psi K_{\rm S}}=0.62$, also the uncertainty due to
$P_c(\nu\overline{\nu})=0.40\pm0.06$ has been shown.}\label{fig:Xpos}
\end{figure}

\begin{figure}
\centerline{\rotate[r]{
\epsfysize=10.3truecm
{\epsffile{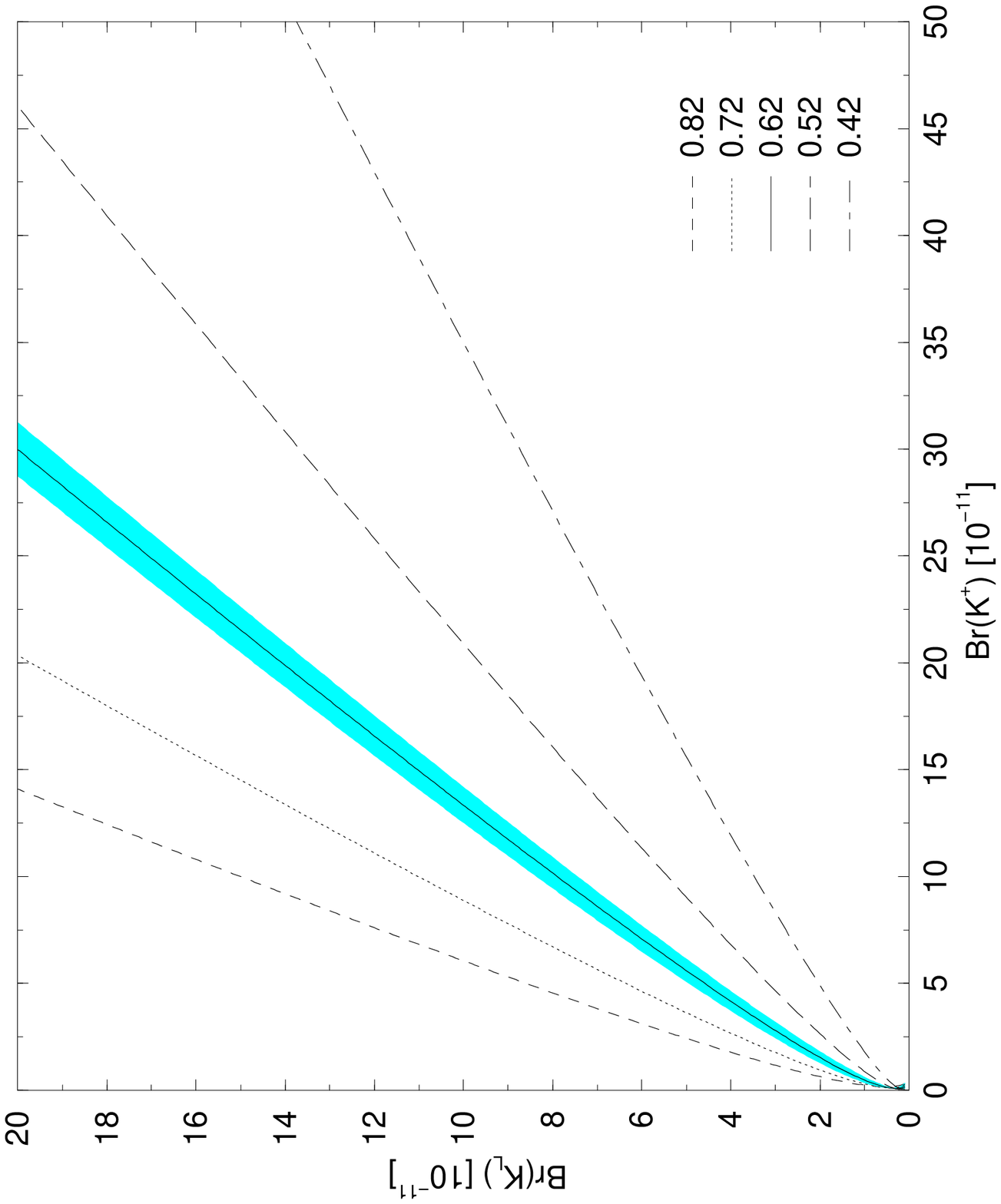}}}}
\caption{$Br(\klnn)$ as a function of $Br(\kpnn)$ for several 
values of $a_{\psi K_{\rm S}}$ in the case of ${\rm sgn}(X)=-1$.
For $a_{\psi K_{\rm S}}=0.62$, also the uncertainty due to
$P_c(\nu\overline{\nu})=0.40\pm0.06$ has been shown.}\label{fig:Xneg}
\end{figure}

A more detailed presentation is given in Figs.~\ref{fig:Xpos} and 
\ref{fig:Xneg}. In Fig.~\ref{fig:Xpos}, we show $Br(\klnn)$ as a function 
of $Br(\kpnn)$ for chosen values of $a_{\psi K_{\rm S}}$ and 
${\rm sgn}(X)=+1$. The corresponding plot for ${\rm sgn}(X)=-1$ is shown 
in Fig.~\ref{fig:Xneg}. It should be emphasized that the plots shown in 
Figs.~\ref{fig:Xpos} and \ref{fig:Xneg} are universal for all MFV models. 
Table~\ref{ANA} and Figs.~\ref{fig:Xpos} and \ref{fig:Xneg} make it clear 
that the measurements of $Br(\klnn)$, $Br(\kpnn)$ and $a_{\psi K_{\rm S}}$ 
will easily allow the distinction between the two signs of $X$.
The uncertainty due to $P_c(\nu\overline{\nu})$ is non-negligible but
it should be decreased with the improved knowledge of the charm-quark
mass.

We would like to emphasize that the upper bound on $Br(\klnn)$ in the last
column of Table~\ref{ANA} is substantially stronger than the 
model-independent bound following from isospin symmetry \cite{NIR96}
\begin{equation}
\label{iso}
Br(\klnn) < 4.4 \cdot Br(\kpnn).
\end{equation}
Indeed, taking the experimental bound  
$Br(\kpnn)\le 5.9\cdot 10^{-10}~(90\%~\mbox{C.L.})$ from AGS E787 
\cite{Adler00}, we find
\begin{equation}\label{KL-bound}
Br(\klnn)_{\rm MFV}\le\left\{\begin{array}{ll}
4.9 \cdot 10^{-10} &  {\rm sgn}(X)=+1\\
7.1 \cdot 10^{-10}  &  {\rm sgn}(X)=-1.
\end{array}\right.
\end{equation}
This should be compared with $Br(\klnn) < 26 \cdot 
10^{-10}~(90\%~\mbox{C.L.})$ following from (\ref{iso}), and with the present 
upper bound from the KTeV experiment at Fermilab \cite{KTeV00X}, yielding
$Br(\klnn)<5.9 \cdot 10^{-7}$. The corresponding predictions within the
SM read~\cite{ERICE}
\be 
Br(\kpnn)= (7.5\pm 2.9)\cdot 10^{-11}~, \quad
Br(\klnn)= (2.6\pm 1.2)\cdot 10^{-11}~.
\ee
As can be seen in Table~\ref{ANA} and in Figs.~\ref{fig:Xpos} 
and \ref{fig:Xneg}, the bounds in (\ref{KL-bound}) will be 
considerably improved when $Br(\kpnn)$ and $a_{\rm \psi K_{\rm S}}$ will 
be known better. The experimental outlook for both decays has recently been 
reviewed by Littenberg \cite{LITT00}. The existing measurement \cite{Adler00}
\be 
Br(\kpnn)= \left(1.5^{+3.4}_{-1.2}\right)\cdot 10^{-10}
\ee
should be considerably improved already this year.

\subsection{An Upper Bound on \boldmath{$Br(\klnn)$} from
\boldmath{$Br(B\to X_s\nu\overline{\nu})$}}
The branching ratio for the inclusive rare decay $B\to X_s\nu\overline{\nu}$ 
can be written in the MFV models as follows \cite{ERICE}:
\be\label{BXS}
Br(B\to X_s\nu\overline{\nu})=1.57\cdot 10^{-5}
\left[\frac{Br(B\to X_c e\overline{\nu})}{0.104}\right]
\left|\frac{V_{ts}}{V_{cb}}\right|^2
\left[\frac{0.54}{f(z)}\right]~X^2,
\ee
where $f(z)=0.54\pm0.04$ is the phase-space factor for 
$B\to X_c e\overline{\nu}$ with $z=m_c^2/m_b^2$, and 
$Br(B\to X_c e\overline{\nu})=0.104\pm 0.004$.

Formulae (\ref{bklpn}) and (\ref{BXS}) imply an interesting relation 
between the decays $\klnn$ and $B\to X_s\nu\overline{\nu}$:
\be\label{KLBXS}
Br(\klnn)=42.3\cdot (\imlt)^2
\left[\frac{0.104}{Br(B\to X_c e\overline{\nu})}\right]
\left|\frac{V_{cb}}{V_{ts}}\right|^2
\left[\frac{f(z)}{0.54}\right] Br(B\to X_s\nu\overline{\nu}),
\ee
which is valid in all MFV models. Equation (\ref{KLBXS}) constitutes 
still another connection between $K$- and $B$-meson decays, in addition to 
those discussed already in this paper and in \cite{BBSIN,BB98,NIR,BePe,BePe0}.

Now, the experimental upper bound on $Br(B\to X_s\nu\overline{\nu})$ reads 
\cite{ALPH}
\be\label{BBOUND}
Br(B\to X_s\nu\overline{\nu})<6.4 \cdot 10^{-4}\quad (90\%~\mbox{C.L.}).
\ee
Using this bound and setting $\imlt=1.74\cdot 10^{-4}$ (see (\ref{uimlt})), 
$|V_{ts}|=\vcb$, $f(z)=0.58$ and $Br(B\to X_c e\overline{\nu})=0.10$, we 
find from (\ref{KLBXS}) the upper bound
\be
Br(\klnn)<9.2\cdot 10^{-10}\quad (90\%~\mbox{C.L.}),
\ee
which is not much weaker than the bound in (\ref{KL-bound}). 
As the bound in (\ref{BBOUND}) should be improved in the $B$-factory era, 
also the latter bound should be improved in the next years.

\begin{figure}
\centerline{\rotate[r]{
\epsfysize=10.3truecm
{\epsffile{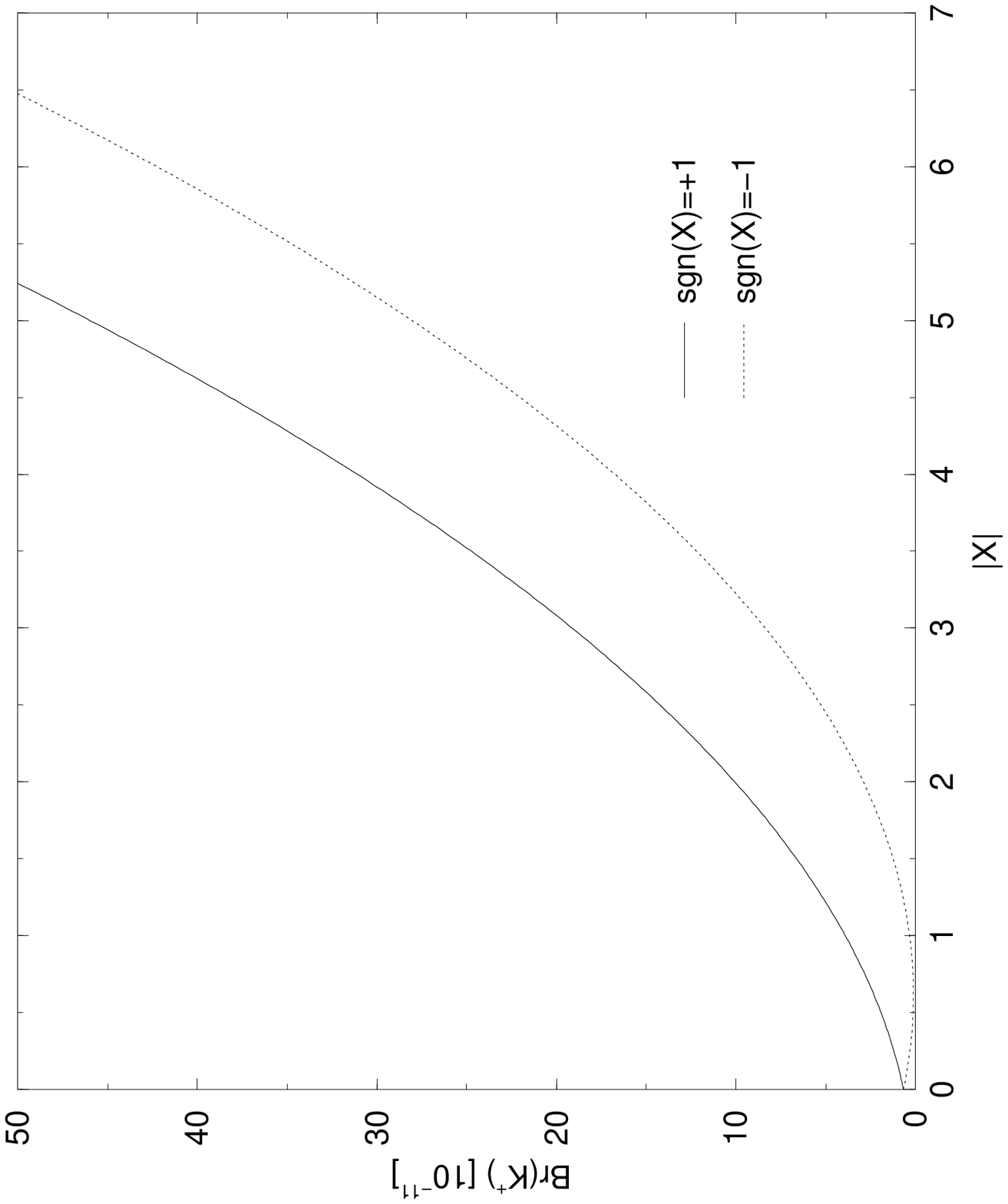}}}}
\caption{$Br(\kpnn)$ as a function of $|X|$ for ${\rm sgn}(X)=\pm 1$ in
the case of $A=0.83$, $(\bar\varrho,\bar\eta)=(0.23,0.35)$
and $P_c(\nu\overline{\nu})=0.40$.}\label{fig:Brp}
\end{figure}

\subsection{Determination of {\boldmath$X$\unboldmath}}
The knowledge of the function $X$ would be a very important 
information, providing constraints on the MFV models. In the SM, 
we have $X\approx 1.5$. Present bounds on the function $X$ from 
$\kpnn$ and $ B\to X_s \nu\overline{\nu}$ within MFV models were recently 
discussed in \cite{BePe}. In particular, from (\ref{BXS}) and 
(\ref{BBOUND}) we find
\be
|X|<6.8,
\ee
which agrees well with \cite{BePe}.

In the future, a theoretically clean determination of $X$ 
will be made possible by determining $\bar\eta$ and $\bar\varrho$ by means 
of $\Delta M_s/\Delta M_d$ and $a_{\psi K_{\rm S}}$ (see (\ref{Rt}) and 
(\ref{rho-eta})), and inserting them into (\ref{RE-IM}) and (\ref{bkpn}).
In this manner, we may calculate $Br(\kpnn)$ as a function of $X$.
The measurement of this branching ratio yields then two values of $|X|$, 
corresponding to ${\rm sgn}(X)=\pm 1$. We illustrate this in 
Fig.~\ref{fig:Brp}, where we plot $Br(\kpnn)$ as a function of $|X|$ for 
${\rm sgn}(X)=\pm 1$. Here we have assumed, as 
an example, $A=0.83$, $(\bar\varrho,\bar\eta)=(0.23,0.35)$, which 
corresponds to $a_{\psi K_{\rm S}}=0.75$, and $P_c(\nu\overline{\nu})=0.40$. 
As expected, $Br(\kpnn)$ is substantially smaller in the case of a negative 
$X$.

Direct access to $|X|$ will also be provided by $Br(\klnn)$, as can 
be seen from (\ref{bklpn}). If a MFV model is realized 
in nature, both determinations have to give the same value of $|X|$. 
This requirement allows us to distinguish between the two branches in
Fig.~\ref{fig:Brp}, thereby offering another way to fix the sign of $X$. 

However, the strategy presented in Subsection~\ref{sec:GOLD}, which is based 
on Figs.~\ref{fig:Xpos} and \ref{fig:Xneg} and involves just 
$a_{\psi K_{\rm S}}$, $Br(\kpnn)$ and $Br(\klnn)$, is much more elegant to 
check whether a MFV model is realized in the $K\to\pi\nu\overline{\nu}$ 
system and -- if so -- to determine ${\rm sgn}(X)$. In order to determine 
also $|X|$, $\Delta M_s/\Delta M_d$ is needed as an additional input,
as we have seen above.

\section{Conclusions}
In this paper, we have explored the determination of $\sin 2\beta$ 
through the standard analysis of the unitarity triangle, the CP asymmetry 
$a_{\psi K_{\rm S}}$, and the decays $K\to \pi\nu\overline{\nu}$ in MFV
models, admitting new-physics contributions that reverse the sign of the 
corresponding generalized Inami--Lim functions $F_{tt}$ and $X$. Our
findings can be summarized as follows:
\begin{itemize}
\item There are bounds on $\sin 2\beta$, which can be translated into 
lower bounds on $a_{\psi K_{\rm S}}$. For $F_{tt}>0$, 
$(a_{\psi K_{\rm S}})_{\rm min}=0.42$ \cite{ABRB}, whereas we obtain a 
stronger bound of $(a_{\psi K_{\rm S}})_{\rm min}=0.69$ in the case of 
$F_{tt}<0$. Consequently, for $0.42<a_{\psi K_{\rm S}}<0.69$, the full 
class of MFV models with $F_{tt}<0$ would be excluded; for
$a_{\psi K_{\rm S}}<0.42$, even all MFV models would be ruled out. 
The reduction of the uncertainties of the relevant input parameters 
could improve these bounds in the future. We have also discussed 
strategies to determine the sign of $F_{tt}$ directly, allowing 
interesting consistency checks of the MFV models. 

\item The most recent $B$-factory data are no longer in favour of small 
values of $a_{\psi K_{\rm S}}$, and the present world average of 
$0.79\pm0.10$ does not even allow us to exclude the case corresponding to 
$F_{tt}<0$. Consequently, an important role may be played in the future by 
the upper bound on $a_{\psi K_{\rm S}}$ that is implied by $|V_{ub}/V_{cb}|$. 
Since the BaBar and Belle results are not fully consistent with each other,
the measurement of $a_{\psi K_{\rm S}}$ will remain a very exciting
issue. Let us hope that the situation will be clarified soon.

\item We have generalized the SM analysis of the unitarity 
triangle through $K\to\pi\nu\overline{\nu}$ to MFV models, allowing
negative values of $X$. In particular, we have explored the behaviour 
of $Br(\klnn)$ as a function of $a_{\psi K_{\rm S}}$ and $Br(\kpnn)$ 
for the general MFV model. This is an important excercise, since the 
latter two quantities will be known rather precisely before $Br(\klnn)$ 
will be accessible. In this context, we have pointed out that for given 
$Br(\kpnn)$ and $a_{\psi K_{\rm S}}$, only two values for $Br(\klnn)$ are 
possible in the full class of MFV models, which correspond just to the
two signs of $X$ and are independent of any new parameters present in 
these models. Consequently, the measurement of this branching ratio will 
either select one particular class of MFV models, or will exclude 
all of them.

\item At present, the existing lower and upper bounds on 
$a_{\rm \psi K_{\rm S}}$ in the MFV models allow us to find 
{\it absolute} lower and upper bounds on the branching ratio 
$Br(\klnn)$ as a function of $Br(\kpnn)$. We find that the present 
upper bounds on $Br(\kpnn)$ and $\vub$ imply an absolute upper 
bound $Br(\klnn)< 7.1 \cdot 10^{-10}$ ($90\%$ C.L.), which is 
substantially stronger than the bound following from isospin symmetry.
On the other hand, the experimental upper bound on 
$Br(B\to X_s\nu\overline{\nu})$ implies 
$Br(\klnn)<9.2\cdot 10^{-10}$ ($90\%$ C.L.).
\end{itemize}
The present paper, in conjunction with earlier analyses 
\cite{UUT,ABRB,ERICE,BePe}, demonstrates the simplicity of the MFV models, 
allowing transparent and general tests of these models without
the necessity to assume particular values for their new parameters. 

It will be exciting to follow the development in the experimental 
values of $a_{\rm \psi K_{\rm S}}$, $Br(\kpnn)$, $Br(\klnn)$, 
$Br(B\to X_s\nu\overline{\nu})$ and $\Delta M_s/\Delta M_d$. Possibly already 
before the LHC era we will know whether any of the MFV models survives all 
tests discussed here and in \cite{BBSIN,NIR,UUT,ABRB,ERICE,BePe}, or whether 
new operators and/or new complex phases are required to describe the data.

\section*{Acknowledgements}
A.J.B. would like to thank Piotr Chankowski and {\L}ucja S{\l}awianowska 
for discussions.
This work has been supported in part by the German Bundesministerium 
f\"ur Bildung and Forschung under the contract 05HT9WOA0.

\vfill\eject

\end{document}